\documentclass{kapproc} % Computer Modern font calls

\usepackage{procps} 

\usepackage[dvips]{graphicx}

\upperandlowercase

\kluwerbib % will produce this kind of bibliography entry:

\begin{document}

%------------ article title  ------------------->>

% If you use \\'s , please supply an alternate version of the title
% in square brackets, i.e., 
%\articletitle[Communism, Sparta, and Plato]
%{COMMUNISM, SPARTA,\\ and PLATO}

\articletitle[Continuous Star Clusters formation in the spiral NGC45]{CONTINUOUS  STAR CLUSTER FORMATION IN THE SPIRAL NGC 45}

%% optional, to supply a shorter version of the title for the running head:
%%\chaptitlerunninghead{}

%\author{}

%% multiple authors may be separated with \\
%% \author{Samuel Bostaph\\
%% and Gregor Kariotis}

%------ author/affiliation choices -------------->>

%% Single author

% \author{}
% \affil{}
% \email{}

%% Multiple authors, single affiliation

% \author{Marcelo Mora-Genskowsky} \author{Soeren S. Larsen}
%  \author{Markus kissler-Patig}                      % \and   % <=== Type in \and before the last author so that `and' will
% \author{Markus kissler-Patig}    % print between the last two authors
                           % in the table of contents.

% \affil{}

%% Multiple authors, multiple affiliations

% First method:
%--------------

%\author{}
%\affil{}
%
%\author{}
%\affil{]
%
%etc.%

% Second method:
%--------------

%\author{author\altaffilmark{1}, author\altaffilmark{2}, 
%         author\altaffilmark{1,3}}%
%
%\altaffiltext{1}{Affiliation}
%\altaffiltext{2}{Affiliation}
%\altaffiltext{3}{Affiliation}

% Third method:
%--------------

\author{Marcelo Mora\altaffilmark{1}, S$\o$ren S. Larsen\altaffilmark{2} and 
         Markus Kissler-Patig\altaffilmark{1}}\

\altaffilmark{1} {ESO Garching, Germany}; \
\altaffilmark{2}{ESO / ST-ECF, Germany} 
%\altaffilmark{3}Second affiliation}

%------ prologue, abstract, keywords ----------->>
% optional prologue
%\prologue{<text>}{<author, year>}

% optional abstract
 \begin{abstract}
We determined ages for 52 star clusters  with masses $\le 10^6$ $M_{\odot}$ in the low surface brightness spiral galaxy NGC 45. 
Four of these candidates are old globular clusters located in the bulge. The remaining ones span  a large age range. 
The cluster ages suggest a continuous star/cluster formation history without evidence for bursts, 
consistent with the galaxy being located in a relatively  unperturbed environment in the outskirts of the Sculptor group.

%The cluster ages show a  galaxy  with continuous star/cluster formation despite being in an unperturbed environment.

 \end{abstract}

% optional keywords
% \begin{keywords}
   % Text, text...
% \end{keywords}

%------------ body of article ------------------->>
\section{Introduction}
Star clusters are an ideal tool to study star formation histories in nearby 
galaxies. Historically, star clusters have been studied in the Milky Way, Local 
Group, ellipticals, mergers and starbursts. But there are only  few studies in 
low luminosity galaxies. The closest examples with similar luminosity to NGC 45
are LMC, SMC, and M33 (e.g.~C$\hat{o}$t$\acute{e}$ et al.~\cite{cote}).  Star clusters 
in these galaxies show different star formation histories.  LMC shows 3 bursts of star formation 
(Pietrzynski \cite{pietrzynski}) while  SMC shows a more uniform star cluster formation history.
M33 also shows a continuous star cluster formation (Chandar \cite{chandra}).
But beyond the Local Group, the properties of cluster populations and  star formation in general  are poorly known in normal galaxies.
%NGC 45 is an outlying member of the sculptor group. It has a surface brightness $M_B$=-17.13 
%(Bottinelli \cite{bottinelli}).
NGC 45 is an interesting galaxy: It is an outlying member of the sculptor group (m-M)=28.42 $\pm0.41$ (Freedman et al.~\cite{freedman}).   
It is located in an unperturbed environment and it has a low luminosity of $M_B$=-17.13 (Bottinelli \cite{bottinelli}).
In the presented work in progress, we study the star formation history of this spiral galaxy by looking at the NGC45  star cluster system.

\section{Observations}
%NGC 45 is  the faintest member of the sculptor group.It has a similar luminosity 
%to the SMC, LMC and M33(C\^ot\'e et al. \cite{cote}). It is located at ~5 Mpc. from the MW.
%In our study we used  HST ACS and  WFPC2  images trought the filters F336W (~U), 
%F435W (~B), F555W (~V), F814W (~I).
%The object detection was done using SExtractor, an the  aperture photometry
%using IRAF/PHOT. We selected our cluster candidates according to their size.
%We selected objects with extended FWHM according to SExtractor and Baolab Ishape \ref{baolab}  
%Finally all round extended  62 objects were treated as a star clusters candidates.

In our study we used HST ACS and WFPC2 data. We acquired the images using the filters 
F435W ($\sim$B), F555W ($\sim$V), F814W ($\sim$I) and F336W ($\sim$U). The object detection was performed using SExtractor.
The aperture photometry of 6 pixels radius was done using the IRAF/PHOT task. We selected our cluster candidates
by looking at the physical object sizes. For this purpose we used two criteria. We selected cluster candidates as 
objects with extended FWHM according to SExtractor and as well as with extended FWHM according to BAOLAB ISHAPE (Larsen \cite{baolab}). 
%Finally we considered as cluster candidates those
%objects who shows extended FWHM in both criteria. 
52  ``round'', extended objects satisfied the criteria and 
were treated as star clusters.

%
% Finally the cluster candidate selection was performed 
%using SExtractor\ref{sextractor} and Baolab Ishape \ref{baolab} size object criteria.
%Objects that shows extended FWHM were considered has a clusters candidates. Finally
%all round, extended objects were treated as star clusters candidates.   

\section{Color Magnitude diagram}

The color-magnitude diagram (Figure 1) shows two main cluster 
populations:
a red one around $F435W-F555W \sim 0.7$ (likely globular clusters) and 
another, broader distribution around $F435W-F555W \sim 0$ which are likely young cluster candidates.

%The Color Magnitude diagram (Figure 1) shows two colors cluster populations:
%a red one around $F435W-F555W \sim 0.7$ (globular clusters) and 
%a blue one around $F435W-F555W \sim 0$. 
  %which corresponds to a young population.
%(globular clusters) and an intermediate one around $F435W-F555W \sim 0.5$. 

%The Color Magnitude diagram (figure 1), shows three different cluster population:
%a blue one ( around $F435W-F555W \sim 0$) and a red one ( $F435W-F555W\sim0.7$). 
%The first population( filled square)  correspond to the young population and the second one
%(filled circles) correspond to  the old population (i.e. 3 globular clusters). Also and 
%intermediate population is shown with filled triangles which is derived from the the figure 2

% all  NGC45 objetcs are shown. Two color 
%distribution of clusters can be %
%
%showing that the star clusters are distributed in to colors  
%dots correspond to all NGC 45 stars. Clusters
%All the cluster candidates are shown in filled squares, triangles and dot.

\begin{figure}[h!]
\centering
\includegraphics[width=8cm]{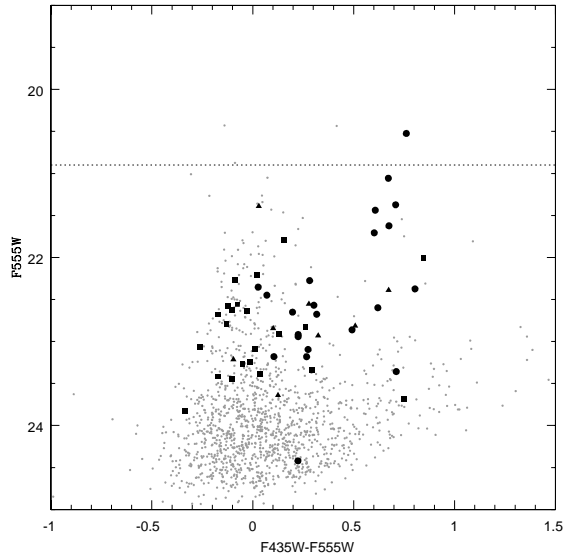}
%\caption{\protect\inx{Oscillograph} for memory address access ....memory plane.}
\label{colormagnitude}
\caption{ NGC 45 Color Magnitude Diagram. Here we plot objects with $UBVI$ photometry. Approximately 3400 stars 
are shown  (black dots) and 52 star cluster candidates. 
Filled squares correspond to a  1-10 Myr cluster population, filled triangles to cluster age 10-100 Myr and
filled circles to cluster age 100-1000 Myr (for age derivation see below). 
The dashed line is the TO of the old MW globular cluster system $M_{VTO}\sim -7.4$}.

%NGC 45 Color Magnitude Diagram. Here we plot $\sim$ 8000 stars
%(black dot) and our 62 cluster candidates in differents filled points, which also
%means differents ages: Filled squares correspond to a  1-10 Myr clusters population, filled triangles 10-100 Myr and
%filled circles 100-1000 Myr.(for age derivation see below).  }
%The filled dots are the cluster candidates. The black dots are NGC45 stars. The plot 
%reveals tree different populations of cluster candidates. A blue one ( around F435W-F555W~0 ), 
%a red one ( F435W-F555W ~0.7)  and a green one. The first one correspond to a young population,
% the second correspond to an intermediate age population, and the third one correspond to an 
%old population (i.e. 3 globular clusters). }
\end{figure}

\section{Color-Color diagram: Ages and Masses}

%One of the principal problems in the star clusters studies is the fact that we do not know the
%extinction of the clusters,  ages and  masses.
%Bik et al. \cite{arjan} proposed a method known as the 3D fitting method. This consists in minimizing the extinction, 
%mass and age for each single cluster using a SSP model (in our case :GALEV \cite{galev}).
One of the principle problems in deriving ages, metallicities and masses for star clusters in spirals, is the fact that 
we do not know the extinction towards the individual objects.
Bik et al. (\cite{arjan}) proposed a method known as the 3D fitting method to  solve this problem.
It consists in minimizing the extinction, mass and age for each single cluster using a SSP model 
(In our case: GALEV, Ander et al.~\cite{galev}) assuming a fixed metallicity. We apply this method to our dataset assuming a 
solar metallicity for all clusters.

%age and mass derivation.
%In order to derives ages an mass we need to know some how  the cluster reddening. In order to
%obtain this we used the 3D fitting method (Bik et al. \cite{arjan}) which minimize reddening, 
%mass and age for each single candidate using a SSP model (in our case :GALEV \cite{galev} ).

\begin{figure}[h]
\centering
\label{figuras}
\includegraphics[width=5.6cm]{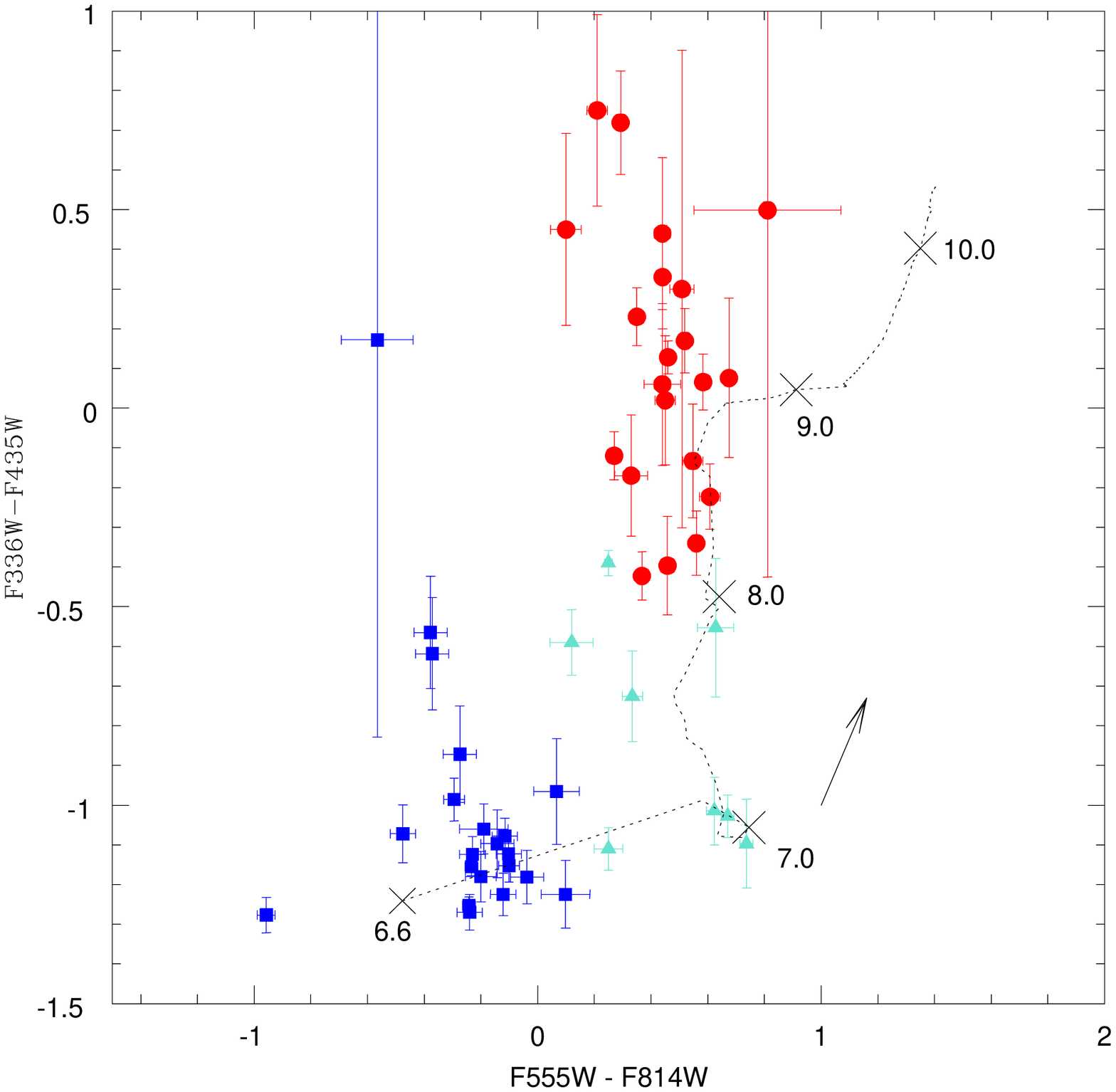}
\includegraphics[width=5.6cm]{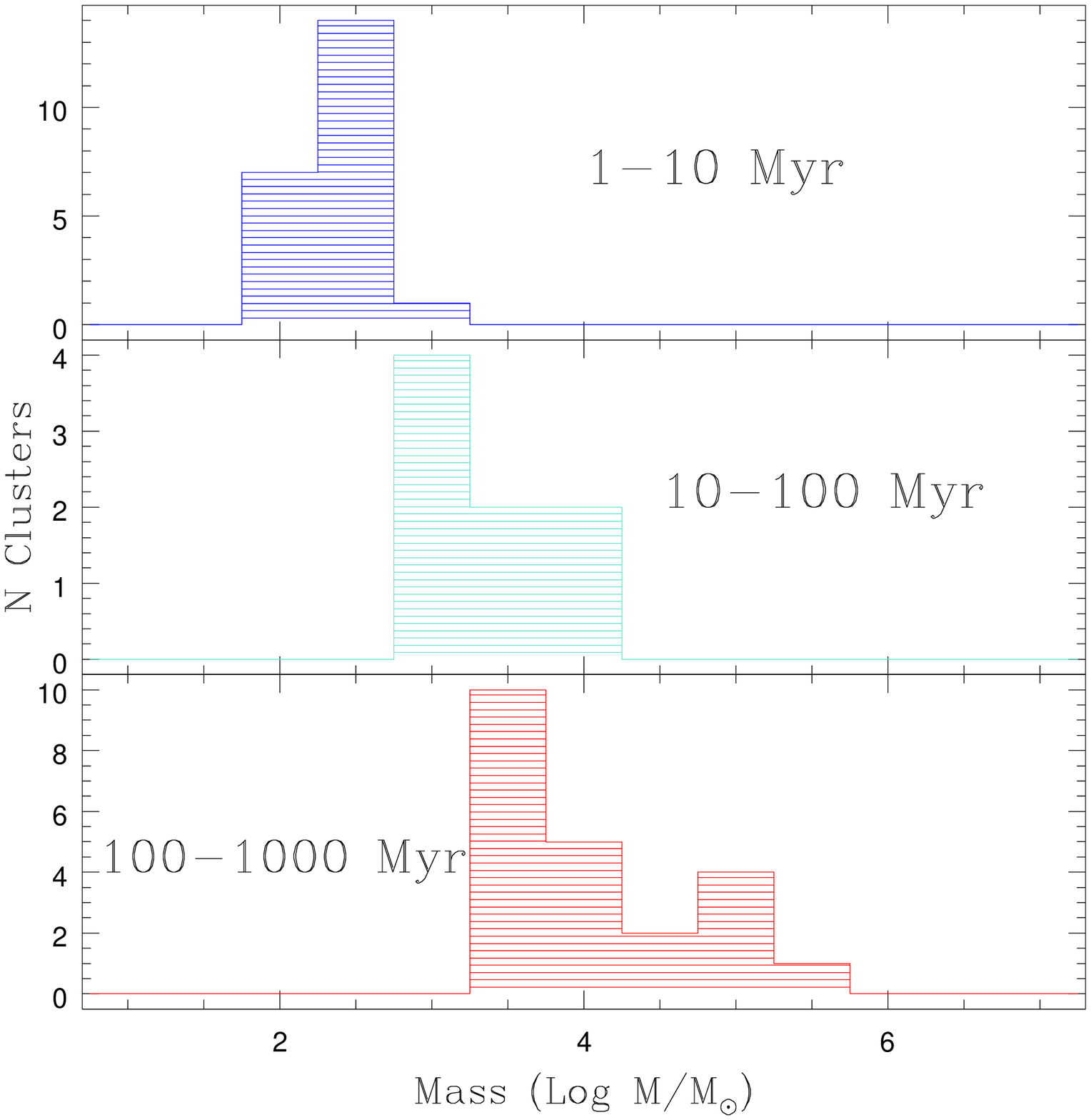}
\caption{Left: color-color diagram of the cluster candidates . Filled circles are  clusters
  with ages 100 Myr -1 Gyr. Triangles  are the clusters with intermediate ages: 10
 to 100 Myr. And the third group (filled squares) is younger than 10 Myr. 
 The dashed line is the SPP GALEV model for a salpeter IMF and a solar metallicity  (Z=0.02),
 for ages 1 Myr  to 10 Gyr (The crosses indicates the log of the age).
 Right: Mass distribution as a function of age. Each bin corresponds to 0.5 Log $M/M_{\odot}$.}

\end{figure}

The color-color diagram in  figure 2 (left plot) shows all the star clusters corrected for reddening.
We use the same symbols as in  Figure 1.
The arrow indicates the reddening corresponding to a 1 magnitude extinction in $V$.
A theoretical SSP model track for ages between 1 Myr and 10 Gyr and a solar metallicity is also shown in dashed line. 
%The clusters are distributed across the theoretical track and show a continuous star formation.
The clusters are distributed across the theoretical track and show an age distribution 
without signatures of discrete bursts.
The estimated masses of the clusters are shown in the right part of the figure 2 for three age bins.%as a function of age.
The upper part shows 22 clusters  between 1 - 10 Myr old, 
the middle part 8 clusters between  10-100 Myr and  the lower part shows 22 clusters between 100-1000 Myr.
Before drawing conclusions about the mass distributions, it is necesary to take size-of-sample effects 
into account ( Larsen \cite{soeren}, Whitmore \cite{whitmore}). Due to the small number 
of clusters in each age  bin, the likelihood of sampling the cluster mass distributions to 
significantly higher masses than those observed is low. This makes it 
difficult to tell whether a physical upper mass limit exists. 
Furthermore, the ages of the clusters in the oldest bin are 
particularly uncertain because of the age-metallicity degeneracy in 
optical broad-band colors. This also translates into an uncertainty on 
the derived masses. Here we have used solar-metallicity SSP models, but 
the colors of the reddest clusters are also consistent with an old, 
metal-poor GC population. A more detailed discussion of these issues 
will be given in Mora et al. (in preparation).
%Therefore,
Considering this, we can say that
%It is necessary to take in account the size of sample effects.
%Due to we have a small statistical sample  it is not very likely that we are observing
%Before to make any assessment, is necessary to take in account  size of sample effects.
%Due to the low sample number it is not very likely probably that we are mapping the mass function
%up to $10^{5}$M$_{\odot}$. Therefore we can not make a strong assessment. Taking this
%in account, we can say:
younger clusters are in general less massive than older ones and there are no massive young clusters.
%Also it is necesary to consider the fact that there is a completeness factor
%it is not very likly that the more massive young cluster is real and it is 
%NOt enough sample means a posibiliti of to find the rare
%si no hay mucha muestra estadistica es poco probable encontrar los objetos raros
%Intermediates and old age clusters shows more massive clusters than young ones.
There are more massive clusters at intermediate and old ages when compared to the youngest bin.
No clusters more masive than $10^{6}$ M$_{\odot}$ are observed.

%
%But there are massive clusters at intermediater  and older ages.

%The Color Color Diagram  of the star clusters (Figure 2) shows a continuous 
%star cluster formation. The  symbols correspond to a different age ranges (same symbols as the figure 1).
%The masses of the clusters are shown in Mass Distribution as function of age plot. The top correspond
%to the blue cluster population, the middle distribution correspond to the intermediate population 
%and the bottom correspond to the old cluster population.

%The Mass distribution as function of age. shows  

%The Color Color diagram  for the cluster candidates (figure 2) shows a continuous 
%star cluster formation and three different clusters population can be observed,
%the different symbols are different age ranges (same symbols as the figure 1). 

%The mass-age distributions shows that older clusters are more massive than younger.
%suggesting that massive clusters also can be form in low surface brightness galaxies.
%Also the figure shows the clusters mass distribution.
%where is posible to apreciate that older clusters are more massive than youngers.

% showing tree
%clusters populations and their mass distribution  acording to the ages.

%The mass-age distribution (Figure 3) shows that old clusters are more massive than the younger 
%ones. The colors correspond to the same population as in the color-color diagram 

\section{Summary and conclusions}
%The color magnitude diagram shows two cluster populations which are in concordance with
%an old (1 Gyr)  and a young population ($\le$10 Myr).

The color magnitude diagram shows two main cluster populations which 
are in concordance with an old ($\ge$ 1 Gyr) globular cluster-like population and
younger objects more similar to the open clusters in the Milky Way. Most
of the latter have young ($\le$ 100 Myr) ages, possibly due to cluster
disruption and fading.

The existence of intermediate-age clusters is deducted from the color-color diagram.
% ($\le$Myr).  An intermediate age population is further detected in the color color plot.
This  shows that NGC 45 is a galaxy with continuous star formation history.

The mass distribution as a function of age shows that more massive clusters are in general older than the less massive.
This is also observed in M33 and also both galaxies shows similar masses 
ranges (between $10^{2}$ M$_{\odot}$ up to 10$^{5}$M$_{\odot}$) (Chandar et al.~\cite{chandra}).

The role of  size-of-sample effects needs to be further investigated.

%The continuous star formation is also observed in  M33 and the LMC but not in the Milky way.
%The cluster ages shows an intermediate population which is not present in the Milky Way.

Finally NGC 45 provides evidence  that  unperturbed low luminosity spiral galaxies can show continuous 
cluster formation.

\end{document}